\documentclass[a4paper,11pt]{article}

\usepackage{pos}

\title{Color-magnetic correlations in SU(N) lattice QCD}
\ShortTitle{Color-magnetic correlations in SU(N) lattice QCD}

\author*{Hideo Suganuma}
\author{Atsuya Tokutake}
\author{Kei Tohme}

\affiliation{Department of Physics, Graduate School of Science, Kyoto University, \\
Kitashirakawa-oiwake, Sakyo, Kyoto 606-8502, Japan}

\emailAdd{suganuma@scphys.kyoto-u.ac.jp}
\emailAdd{tokutake@gauge.scphys.kyoto-u.ac.jp}
\emailAdd{tohme@ruby.scphys.kyoto-u.ac.jp}

\abstract{
Motivated by color-magnetic instabilities in QCD, we investigate field-strength correlations in both SU(2) and SU(3) lattice QCD.
In the Euclidean Landau gauge, 
we numerically calculate 
the perpendicular-type color-magnetic correlation, 
$C_{\perp}(r) \equiv g^2 \langle H_z^a(s)H_z^a(s + r\hat \perp)) \rangle$ with $\perp \equiv x, y$, 
and the parallel-type one,  
$C_{\parallel}(r) \equiv g^2 \langle H_z^a(s)H_z^a(s + r\hat \parallel) \rangle$
with $\parallel~\equiv z, t$. 
In the Landau gauge, all two-point field-strength correlations 
$g^2 \langle G^a_{\mu\nu}(s)G^b_{\alpha\beta}(s')\rangle$
are described by these two quantities, 
due to the Lorentz and global SU($N_c$) color symmetries.
Curiously, the perpendicular-type color-magnetic correlation 
$C_{\perp}(r)$ is found to be always negative for arbitrary $r$, 
except for the same point of $r=0$. 
The parallel-type color-magnetic correlation $C_{\parallel}(r)$
is always positive. 
In the infrared region, 
$C_{\perp}(r)$ and $C_{\parallel}(r)$ 
strongly cancel each other, 
which leads to an approximate cancellation 
for the sum of the field-strength correlations as 
$\sum_{\mu, \nu} \langle G^a_{\mu\nu}(s)G^a_{\mu\nu}(s')\rangle 
\propto C_{\perp}(|s-s'|)+ C_{\parallel}(|s-s'|) 
\simeq 0$.
Next, we decompose the perpendicular-type color-magnetic correlation $C_{\perp}(r)$ 
into quadratic, cubic and quartic terms of the gluon field $A_\mu$.
The quadratic term is always negative, which is explained 
by the Yukawa-type gluon propagator 
$\langle A^a_\mu(s)A^a_\mu(s')\rangle \propto e^{-mr}/r$ with $r\equiv |s-s'|$ 
in the Landau gauge. 
The quartic term gives a relatively small contribution.
In the infrared region, 
the cubic term is positive and tends to cancel with the quadratic term, 
resulting in a small value of $C_{\perp}(r)$.
}

\FullConference{
The XVIth Quark Confinement and 
the Hadron Spectrum Conference 
(QCHSC24), 
19-24 August, 2024, Cairns Convention Centre, Cairns, Queensland, Australia
}

\begin{document}

\maketitle

\section{Introduction}

Since 1973, 
quantum chromodynamics (QCD) has been 
established as the fundamental theory of the strong interaction. 
Due to the asymptotic freedom in QCD, 
its coupling decreases with the renormalization scale, 
and perturbative QCD is applicable to the analysis of high-energy hadron reactions. 
At low energies, however, the coupling becomes strong, and QCD exhibits nonperturbative phenomena 
such as color confinement and 
dynamical chiral symmetry breaking \cite{Handbook_2023}.

In particular, due to the asymptotic freedom, 
QCD has a color-magnetic instability, which involves the 
spontaneous emergence of color-magnetic fields \cite{S77}. 
In other words, the system with zero color-magnetic field is energetically unstable. 
The non-zero color-magnetic QCD vacuum is called  
the Savvidy vacuum and/or the Copenhagen vacuum \cite{NO79}.
Actually, the gluon condensate 
$\frac{\alpha_s}{\pi} \langle G_{\mu\nu}^a G^{\mu\nu}_a\rangle$ 
is positive in the Minkowski space, implying significant excess of 
color-magnetic fields rather than color-electric fields. 
In the QCD vacuum, to recover the rotational symmetry, the color-magnetic systems are to form a fluctuating stochastic domain structure at a large scale \cite{NO79}. 

Considering the fluctuating color fields in the QCD vacuum, Dosch and Simonov proposed the ``stochastic vacuum model" for gauge-invariant field-strength correlators and showed that its infrared exponential damping leads to an asymptotic linear potential \cite{D87S88,DS88}. 
Later, Di~Giacomo~et~al. \cite{GP92,EGM97} and Bali, Brambilla and Vairo \cite{BBV98} found that the gauge-invariant field-strength correlator exhibits infrared exponential damping in lattice QCD.

Motivated by these studies, we 
study the field-strength correlation and its overall behavior in SU(2) and SU(3) lattice QCD \cite{TTS25}. 
In this study, using lattice QCD, we mainly investigate the color-magnetic correlation in the Landau gauge, which has many advantages in terms of symmetries and minimal gauge-field fluctuations. 

\section{Color-magnetic correlations in the Landau gauge}

In Euclidean QCD, the Landau gauge has a global definition to minimize 
\begin{eqnarray}
R[A_\mu^a] \equiv 
\int d^4x~
\{A^a_\mu(x)A^a_\mu(x)\}
\end{eqnarray}
by the gauge transformation. 
In the global definition, the Landau gauge has a clear physical interpretation that it strongly  suppresses in total artificial gauge-field fluctuations associated with the gauge degrees of freedom \cite{ISI09}.

Considering the above-mentioned 
nontrivial color-magnetic structure in the QCD vacuum, 
we investigate the following type of color-magnetic correlation 
in the Landau gauge in lattice QCD:
\begin{enumerate}
\item 
Perpendicular-type color-magnetic correlation 
$C_{\perp}(r) \equiv g^2 \langle H^a_z(s)H^a_z(s+r \hat \perp)\rangle$
\newline
($\hat \perp$: unit vector on the $xy$-plane), 
\item 
Parallel-type color-magnetic correlation 
$C_\parallel(r) \equiv g^2 \langle H^a_z(s)H^a_z(s+r \hat \parallel)\rangle$
\newline
($\hat \parallel$: unit vector  on the $zt$-plane).
\end{enumerate}
In the Euclidean metric, 
one finds 
$\langle H^a_z(s)H^a_z(s+r\hat z)\rangle =\langle H^a_z(s)H^a_z(s+r\hat t)\rangle$
using the four-dimensional rotational invariance.
In the Landau gauge, 
all two-point field-strength correlations  
$g^2 \langle G^a_{\mu\nu}(s)G^b_{\alpha\beta}(s') \rangle$ can be 
expressed with these two correlations 
$C_{\perp}(r)$ and $C_{\parallel}(r)$,
due to the Lorentz and global SU($N_c$) color symmetries. 
%

\section{Lattice QCD setup}

For the nonperturbative analysis 
of the color-magnetic correlation,
we use SU(2) and SU(3) lattice QCD Monte Carlo calculations 
with the standard plaquette action 
at the quenched level.
For the spatial correlation, 
we take both on-axis and 
off-axis lattice data.
On the statistical error of the
lattice data, 
the jackknife error estimate is adopted.

\subsection{SU(3) lattice QCD setup}

For the SU(3) lattice QCD calculations,  
we adopt the following 
four lattices:
\newline
\indent~~~~~
i)~~ $\beta$ = 5.7,~ $L^4=16^4$~~ 
(i.e.,~ $a \simeq$ 0.186 fm,~ 
$La \simeq$ 3.0 fm) 
\newline
\indent~~~~~
ii)~ $\beta$ = 5.8,~ $L^4=16^4$~~
(i.e.,~ $a \simeq$ 0.152 fm,~ 
$La \simeq$ 2.4 fm) 
\newline
\indent~~~~~
iii) $\beta$ = 6.0,~ $L^4=24^4$~~
(i.e.,~ $a \simeq$ 0.104 fm,~ 
$La \simeq$ 2.5 fm)
\newline
\indent~~~~~
iv) $\beta$ = 6.2,~ $L^4=48^4$~~
(i.e.,~ $a \simeq$ 0.0726 fm,~ 
$La \simeq$ 3.48 fm)
\newline
The lattice spacing $a$ is determined so as to reproduce 
the string tension 
$\sigma=0.89~{\rm GeV/fm}$ \cite{ISI09,TSNM02}.
The gauge configurations are picked up with the interval of 1,000 sweeps, 
after the thermalization of 20,000 sweeps. In this study, 
200 gauge configurations are used 
at $\beta$=5.7 and 5.8, and 
800 gauge configurations at $\beta=6.0$. 
For the gluon propagator, we also use 
50 configurations at $\beta$=6.2 \cite{OS12}.

As for the Landau gauge fixing, we use the ordinary iterative maximization algorithm with an over-relaxation parameter of 1.6.
In the Landau gauge, 
since the gluon field is globally minimized, 
we define SU(3) gluon fields 
with the link-variable as
\begin{eqnarray}
{\cal A}_\mu(s) \equiv \frac1{2iag}[U_\mu(s)-U^\dagger_\mu(s)]
-\frac1{2iagN_c}{\rm Tr}[U_\mu(s)-U^\dagger_\mu(s)] \in {\rm su}(N_c) 
\end{eqnarray}
in the fundamental representation. 
This definition is often used 
in the Landau gauge.

\subsection{SU(2) lattice QCD setup}


For the SU(2) lattice QCD calculations, 
we adopt the following three lattices:
\newline
\indent~~~~~
i)~~ $\beta$ = 2.3,~ $L^4=16^4$~~
(i.e.,~ $a \simeq$ 0.18 fm,~~ 
$La \simeq$ 2.9 fm) 
\newline
\indent~~~~~
ii)~ 
$\beta$ = 2.4,~ $L^4=24^4$~~
(i.e.,~ $a \simeq$ 0.127 fm, 
$La \simeq$ 3.0 fm) 
\newline
\indent~~~~~
iii) 
$\beta$ = 2.5,~ $L^4=32^4$~~
(i.e.,~ $a \simeq$ 0.09 fm,~~ 
$La \simeq$ 2.9 fm) 
\newline
The lattice spacing $a$ is determined to reproduce the string tension 
$\sigma=0.89~{\rm GeV/fm}$ \cite{AS99}.
The gauge configurations are picked up with the interval of 200 sweeps, 
after the thermalization of 2,000 sweeps.
We use 400 gauge configurations 
at each $\beta$. 
%
%
The Landau gauge fixing is achieved 
by the ordinary iterative maximization algorithm with over-relaxation parameter of 1.7.
In SU(2) lattice QCD, the gluon field 
$A_\mu(s)$ is directly obtained from the link-variable $U_\mu(s)=e^{iagA_\mu(s)}$ using the general relation of 
$
e^{i\tau^a\theta^a}=\cos\theta+i\tau^a\hat \theta^a \sin\theta 
$
with $\theta \equiv (\theta^a\theta^a)^{1/2}$ and 
$\hat \theta^a \equiv \theta^a/\theta$.

\subsection{Landau-gauge gluon propagator}

\begin{figure}[htbp]
    \centering
    \includegraphics[width=7.5cm]{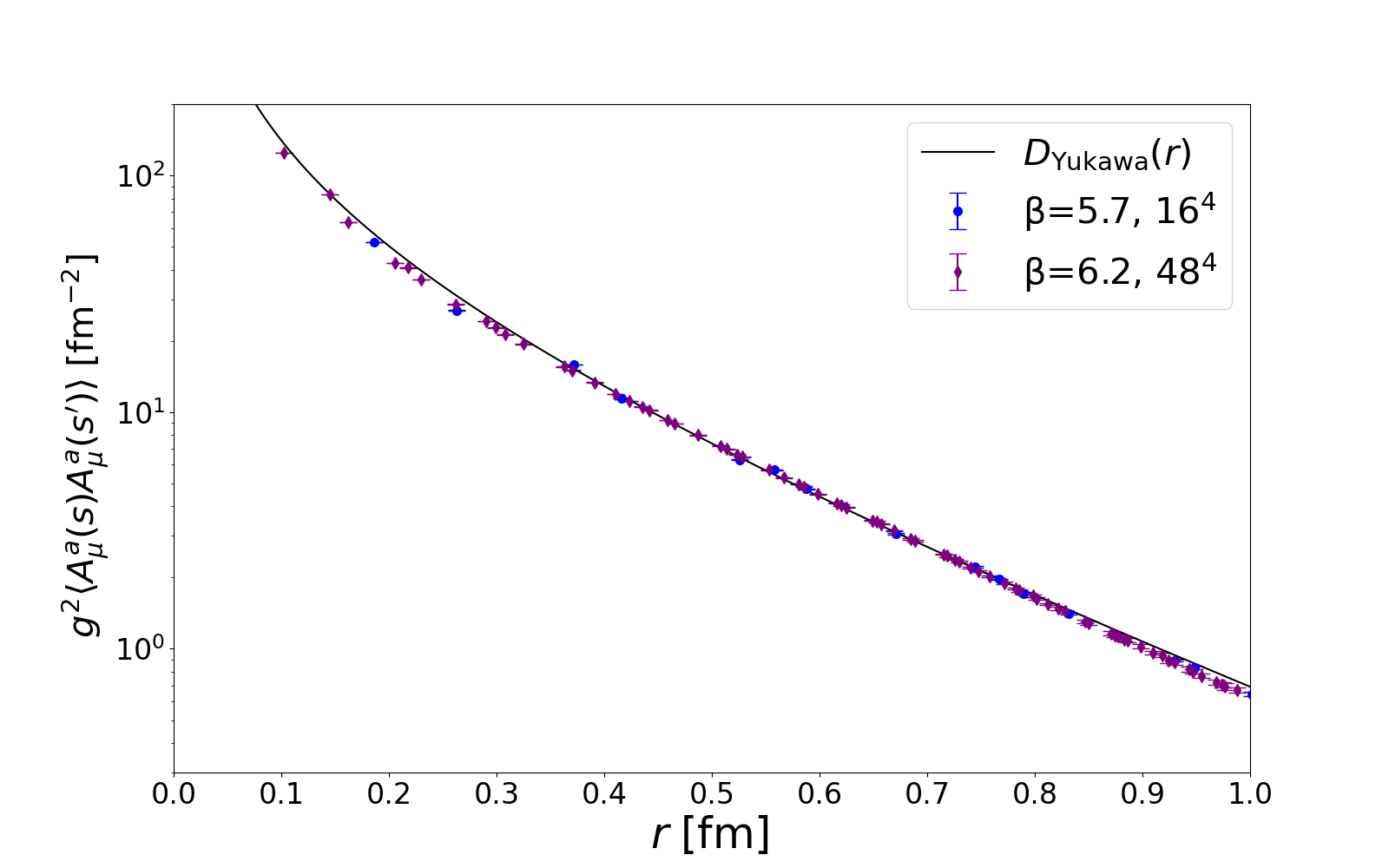}
    \includegraphics[width=7.5cm]{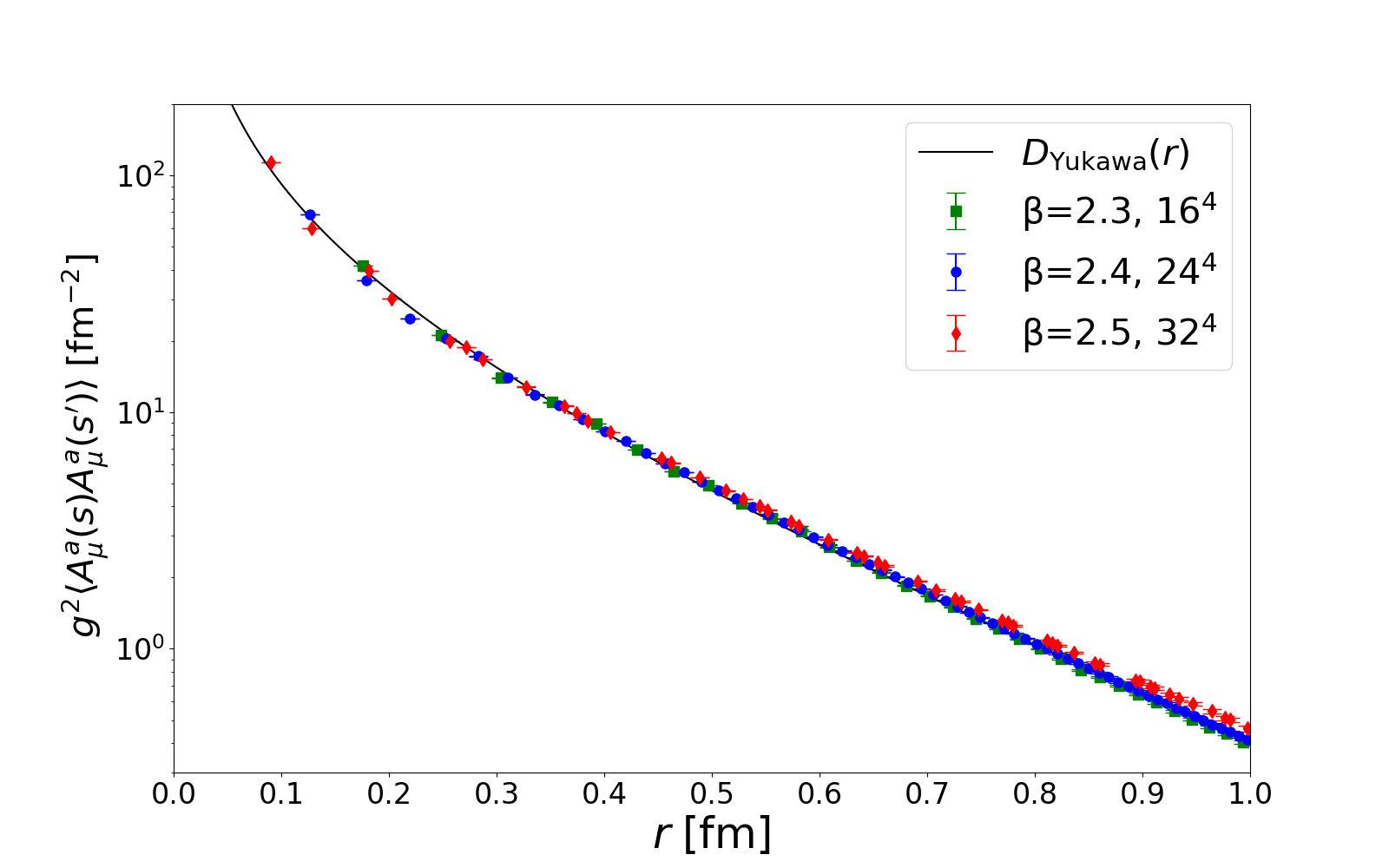}
    \caption{
Landau-gauge gluon propagator 
$D(r)\equiv g^2 \langle A_\mu^a(s)A_\mu^a(s')\rangle$
plotted against $r\equiv |s-s'|$
in SU(3) (left) and SU(2) (right) lattice QCD. 
The curve is the best-fit Yukawa function. 
}
    \label{fig:prop}
\end{figure}

Before proceeding the color-magnetic correlation, 
we examine the Landau-gauge gluon propagator in SU(2) and SU(3) lattice QCD. 
In the Landau gauge, 
all the gluon two-point functions of 
$D_{\mu\nu}^{ab}(s-s')\equiv g^2 \langle A_\mu^a(s)A_\nu^b(s')\rangle$ 
are expressed with the scalar combination 
$g^2 \langle A_\mu^a(s)A_\mu^a(s')\rangle$
\cite{ISI09}, 
which is a single-valued function of 
the four-dimensional Euclidean space-time distance 
$r\equiv |s-s'|$.
%
Figure~\ref{fig:prop} shows the gluon propagator in the Landau gauge 
in lattice QCD.

In both SU(2) and SU(3) QCD, 
the Landau-gauge gluon propagator 
is well described with a Yukawa-type function, 
\begin{eqnarray}
D(r) \equiv
g^2 \langle A_\mu^a(s)A_\mu^a(s')\rangle
\simeq D_{\rm Yukawa}(r),
\quad
D_{\rm Yukawa}(r)\equiv 
A \frac{m}{r}{e^{-mr}}, 
\quad
r\equiv |s-s'|, 
\label{eq:Yukawa}
\end{eqnarray} 
in the wide region of $r = 0.1 - 1.0~{\rm fm}$.
The gluonic mass parameter is estimated as 
$m\simeq 0.660~{\rm GeV}$ for SU(3) QCD and 
$m\simeq 0.676~{\rm GeV}$ for SU(2) QCD.
Since the Yukawa-type gluon propagation 
is natural in the three-dimensional space-time instead of the four-dimensional one, this might relate to some dimensional reduction hidden in nonperturbative QCD \cite{TS24}.

\section{Lattice QCD result for 
color-magnetic correlations} 

In this section, we study the color-magnetic correlations, $C_{\perp}(r)$ and $C_{\parallel}(r)$, in the Landau gauge 
using SU(2) and SU(3) lattice QCD 
at the quenched level.
Note again that 
all the two-point field-strength correlations of 
$g^2 \langle G^a_{\mu\nu}(s)G^b_{\alpha\beta}(s') \rangle$ are expressed with these two correlations. 
In the adopted $\beta$ region, 
$C_{\perp}(r)$ and 
$C_{\parallel}(r)$ at different $\beta$ values  
are found to be approximately single-valued functions 
of $r$ in lattice QCD.

\subsection{Perpendicular-type color-magnetic correlation}

\begin{figure}[htbp]
    \centering
\includegraphics[width=7.5cm]{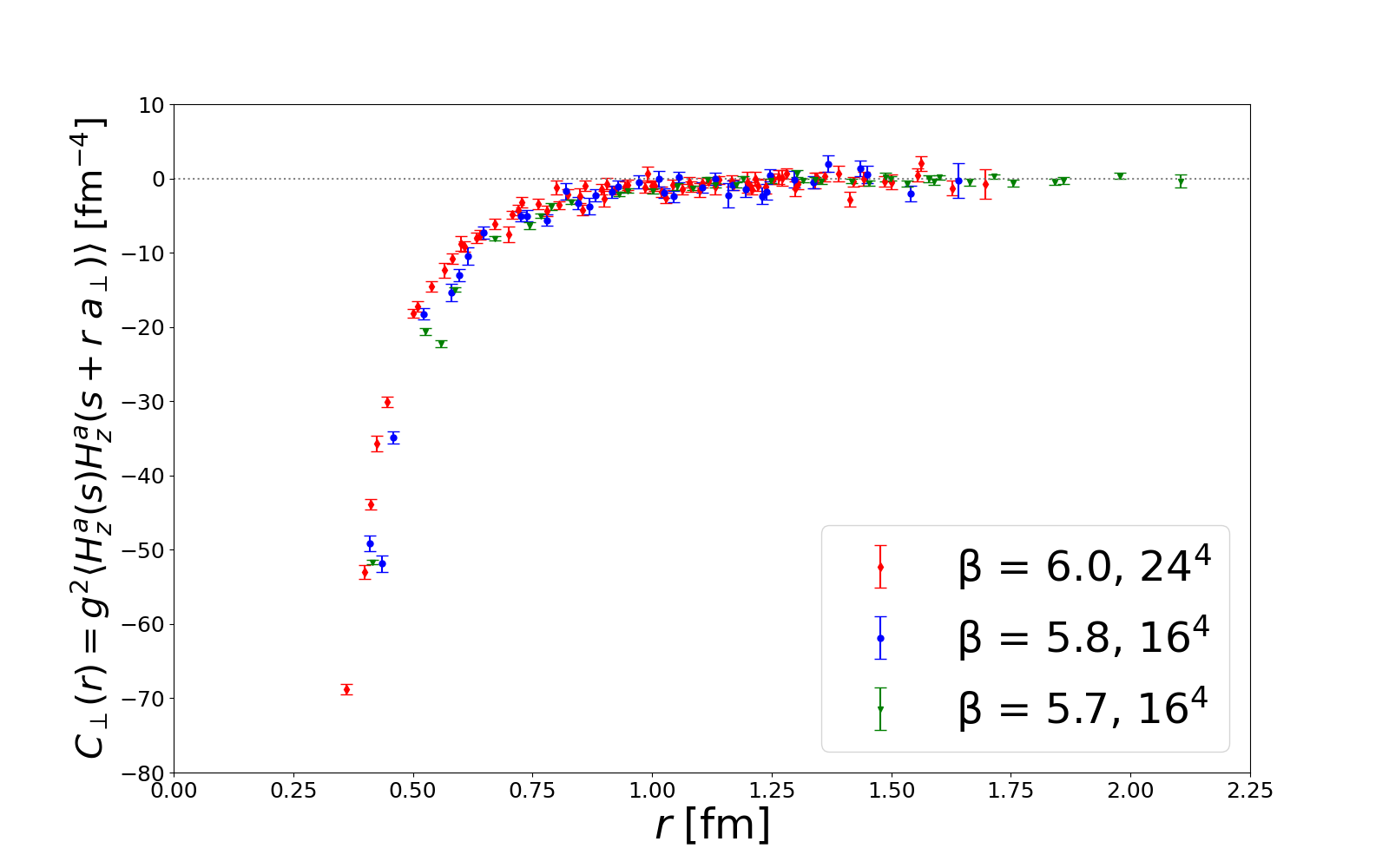}
\includegraphics[width=7.5cm]{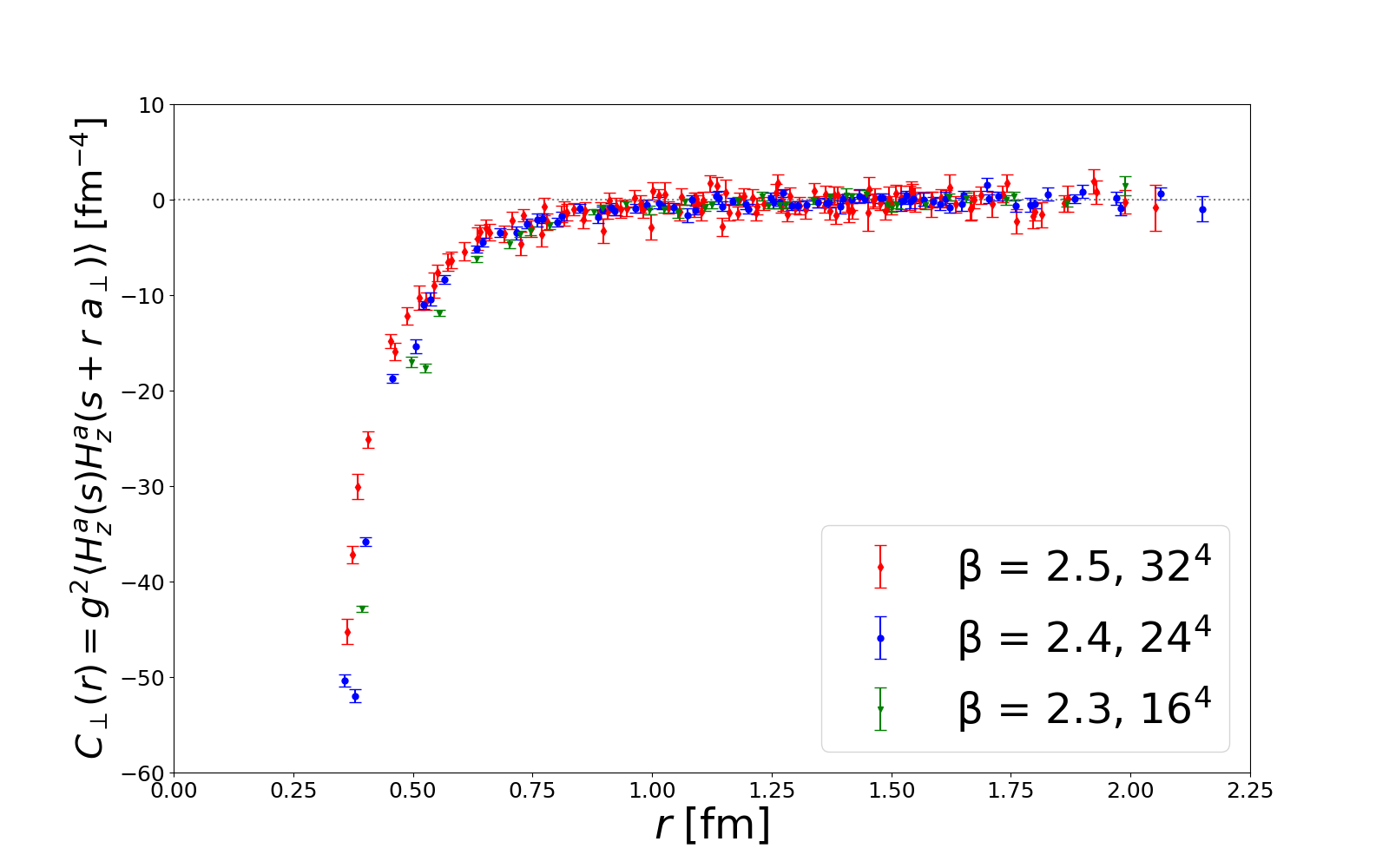}
    \caption{
    The perpendicular-type color-magnetic correlation
$C_{\perp}(r) \equiv g^2 \langle H^a_z(s)H^a_z(s+r\hat \perp)\rangle$ ($\perp~\equiv x,y$)
    in the Landau gauge
    in SU(3) (left) and SU(2) (right) lattice QCD.
    }
    \label{fig:perp}
\end{figure}

To begin with, we investigate the 
perpendicular-type color-magnetic correlation 
\begin{eqnarray}
C_{\perp}(r) \equiv g^2 \langle H^a_z(s)H^a_z(s+r\hat \perp)\rangle 
\quad (\hat \perp: \hbox{unit vector on the $xy$-plane})
\end{eqnarray} 
in the Landau gauge. 
Figure~\ref{fig:perp} shows the numerical result 
in SU(3) and SU(2) lattice QCD.

Curiously, the perpendicular-type color-magnetic correlation is 
{\it always negative}, 
$C_{\perp}(r)<0$, for all values of $r$, 
except for the same point of $r$ = 0.  
In fact, an ``always negative" correlation would be rare in physics, 
whereas ``always positive correlation" and ``alternating correlation"  
have been observed in various areas of physics.

\begin{figure}[htbp]
    \centering
    \includegraphics[width=5cm]{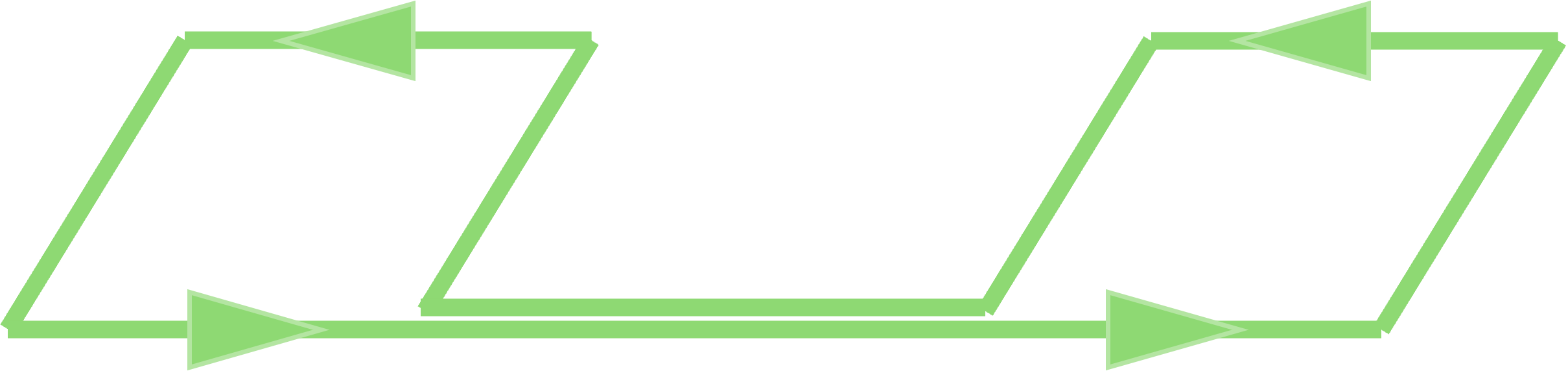}
    \caption{
    An example of the plaquette correlator 
    to extract the gauge-invariant
    field-strength correlation in lattice QCD. 
    }
    \label{fig:plaquette-correlator}
\end{figure}

One might suspect that the gauge fixing has some unphysical effect. 
Then, we also examine gauge-invariant field-strength correlation extracted from the plaquette correlators, as shown in Fig.~\ref{fig:plaquette-correlator}, 
and obtain a similar result. Indeed, the correlation corresponding 
to the perpendicular-type color-magnetic correlation is always negative. 

\subsection{Parallel-type color-magnetic correlation}

\begin{figure}[htbp]
    \centering
    \includegraphics[width=7.5cm]{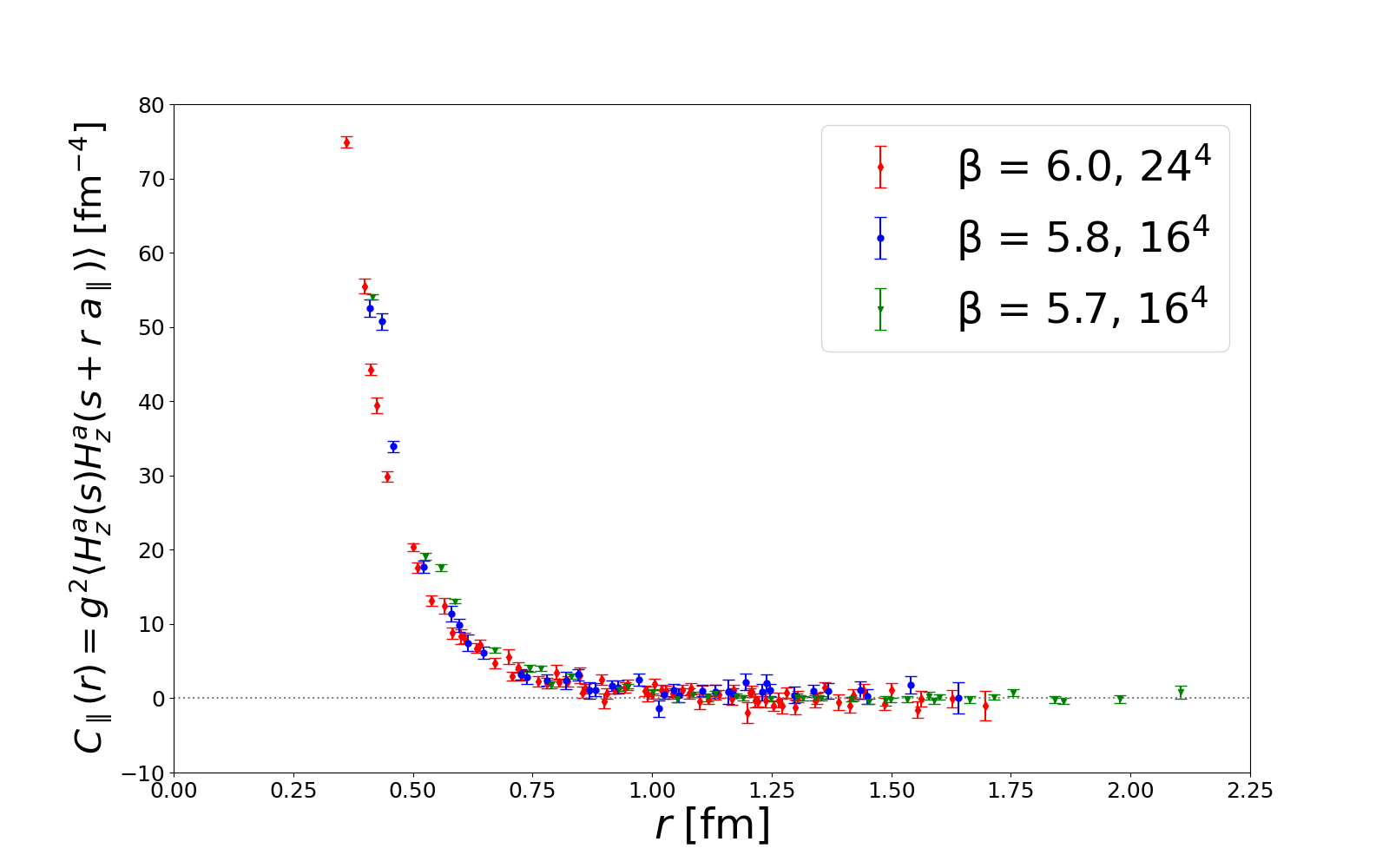}
    \includegraphics[width=7.5cm]{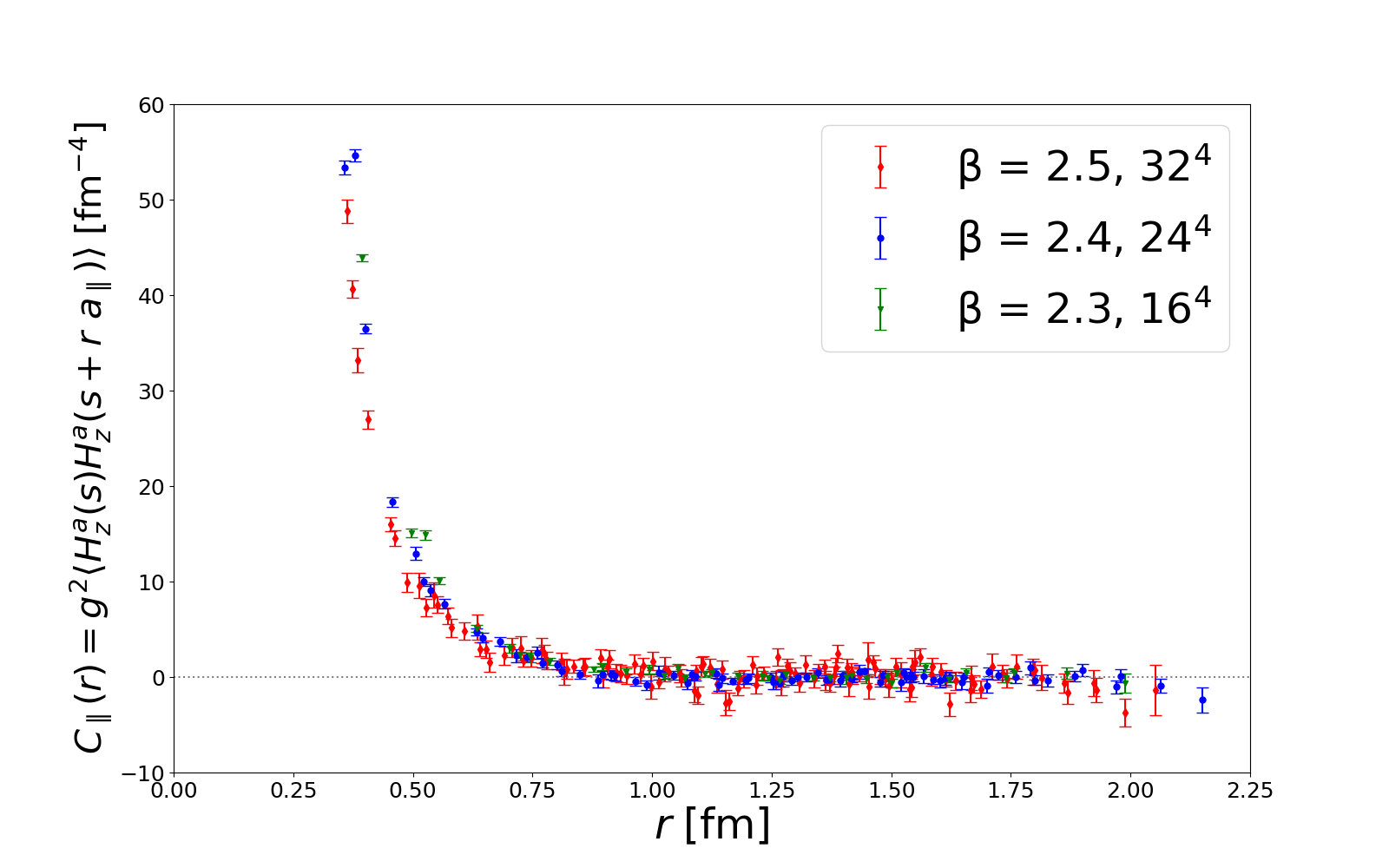}
    \caption{
    The parallel-type color-magnetic correlation
    $C_{\parallel}(r) \equiv g^2 \langle H^a_z(s)H^a_z(s+r\hat \parallel)\rangle$
    ($\parallel~\equiv z,t$)
    in the Landau gauge
    in SU(3) (left) and SU(2) (right) lattice QCD.
    }
    \label{fig:para}
\end{figure}

Next, we show in Fig.~\ref{fig:para} 
the parallel-type 
color-magnetic correlation 
in the Landau gauge,
\begin{eqnarray}
C_\parallel(r) \equiv g^2 \langle H^a_z(s)H^a_z(s+r\hat \parallel)\rangle 
\quad (\hat \parallel: \hbox{unit vector on the $zt$-plane}),
\end{eqnarray} 
in SU(3) and SU(2) lattice QCD.
The parallel-type of $C_{\parallel}(r)$
is found to be always positive, $C_{\parallel}(r) > 0$. 

In the infrared region of $r \gtrsim 0.4~{\rm fm}$, 
the parallel-type color-magnetic correlation $C_{\parallel}(r)$
strongly cancels with 
the perpendicular-type one as 
\begin{eqnarray}
~~~~~~~~~~~~~~ C_{\parallel}(r) \simeq - C_{\perp}(r),
\qquad
{\rm e.g.,~~~} \langle H^a_z(s)H^a_z(s+r\hat z)\rangle 
\simeq 
-\langle H^a_z(s)H^a_z(s+r\hat x)\rangle,  
\end{eqnarray}
which leads to an approximate cancellation 
for the sum of Landau-gauge field-strength correlations,
\begin{eqnarray}
\sum_{\mu, \nu} g^2 \langle G^a_{\mu\nu}(s)G^a_{\mu\nu}(s')\rangle 
=6~[C_{\perp}(r)+C_{\parallel}(r)]
\simeq 0.
\end{eqnarray}

\section{Analysis of color-magnetic correlations in QCD} 

In this section, we try to analyze 
the lattice QCD result 
of the color-magnetic correlation 
in the Landau gauge, particularly considering the origin of 
the negative correlation of $C_{\perp}(r)<0$.

\subsection{Decomposition of the field-strength correlation in terms of the gluon field}

We decompose the field-strength correlation 
$\langle G_{\mu\nu}^a (s) G_{\alpha\beta}^a (s') \rangle$
into three parts, i.e.,  
quadratic, cubic and quartic terms of the gluon field $A_\mu$: 
\begin{eqnarray}
\langle G_{\mu\nu}^a (s) G_{\alpha\beta}^a (s') \rangle
=\langle G_{\mu\nu}^a (s) G_{\alpha\beta}^a (s') \rangle_{\rm quad}
+ \langle G_{\mu\nu}^a (s) G_{\alpha\beta}^a (s') \rangle_{\rm cubic}
+ \langle G_{\mu\nu}^a (s) G_{\alpha\beta}^a (s') \rangle_{\rm quartic}. 
\end{eqnarray} 
Here, the quadratic, cubic and quartic terms are defined by
\begin{eqnarray}
\langle G_{\mu\nu}^a (s) G_{\alpha\beta}^a (s') \rangle_{\rm quad}
&\equiv& \langle  
(\partial_\mu A^a_\nu-\partial_\nu A^a_\mu) (s) 
(\partial_\alpha A^a_\beta-\partial_\beta A^a_\alpha) (s') \rangle,
\cr
\langle G_{\mu\nu}^a (s) G_{\alpha\beta}^a (s') \rangle_{\rm cubic}
&\equiv& 2ig\langle {\rm Tr}~ 
\{(\partial_\mu A_\nu-\partial_\nu A_\mu) (s) 
[A_\alpha, A_\beta] (s') \} \rangle \cr
&+&
2ig\langle {\rm Tr}~
\{[A_\mu, A_\nu] (s) (\partial_\alpha A_\beta-\partial_\beta A_\alpha) (s') 
 \} \rangle, 
\cr
\langle G_{\mu\nu}^a (s) G_{\alpha\beta}^a (s') \rangle_{\rm quartic}
&\equiv& -2g^2\langle {\rm Tr}~ 
\{ [A_\mu, A_\nu] (s) [A_\alpha, A_\beta] (s') \} \rangle,
\end{eqnarray}
where the factor 2 comes from 
${\rm Tr}(T^aT^b)=\frac12\delta^{ab}$.
Among the three terms, the quadratic term can be directly expressed with the gluon propagator $
D^{ab}_{\mu\nu}(s-s')
\equiv
g^2 \langle A^a_\mu(s)A^b_\nu(s') \rangle$ as
\begin{eqnarray}
&&g^2\langle 
G_{\mu\nu}^a (s) G_{\alpha\beta}^b (s') \rangle_{\rm quad} \cr
&=&
\partial_\mu^s \partial_\alpha^{s'}
D_{\nu\beta}^{ab}(s-s')
-\partial_\mu^s \partial_\beta^{s'}
D_{\nu\alpha}^{ab}(s-s')
-\partial_\nu^s \partial_\alpha^{s'}
D_{\mu\beta}^{ab}(s-s')
+\partial_\nu^s \partial_\beta^{s'} 
D_{\mu\alpha}^{ab}(s-s').
\end{eqnarray}
In the Laudau gauge, due to the Lorentz symmetry, 
this quantity can be expressed using  
the scalar combination of the gluon propagator 
$D(r) \equiv g^2 \langle A_\mu^a(s)A_\mu^a(s') \rangle$, 
which is a single-valued function of the four-dimensional Euclidean distance $r=|s-s'|$.

\subsection{Decomposition of perpendicular-type color-magnetic correlation in the Landau gauge}

For the color-magnetic correlation, 
\begin{eqnarray}
\langle H_z^a (s) H_z^a (s') \rangle
&=& \langle 
(\partial_x A^a_y-\partial_y A^a_x) (s) 
(\partial_x A^a_y-\partial_y A^a_x) (s') \rangle \cr
&+&4ig\langle {\rm Tr}~ \{
(\partial_x A_y-\partial_y A_x) (s) 
[A_x, A_y] (s') \} \rangle \cr
&-&2g^2\langle {\rm Tr}~ 
\{[A_x, A_y] (s) [A_x, A_y] (s') \}\rangle,
\end{eqnarray}
the quadratic term can be described 
with the gluon propagator.
Using the Yukawa-type gluon propagator 
$D_{\rm Yukawa}(r)$ in Eq.~(\ref{eq:Yukawa}) in the Landau gauge, 
we find that the
quadratic term in the perpendicular-type color-magnetic
correlation $C_{\perp}(r)$ becomes always negative:  
\begin{eqnarray}
g^2\langle H_z^a(s) H_z^a(s+r\hat \perp)\rangle_{\rm quad}
&=&g^2 \langle  
(\partial_x A^a_y-\partial_y A^a_x) (s) 
(\partial_x A^a_y-\partial_y A^a_x) (s+r\hat \perp) \rangle \cr
&=&-\frac{Am^4}{3}~\frac{e^{-mr}}{mr}\left(1+\frac{1}{mr}+\frac{1}{m^2r^2}\right) < 0 
\quad (\perp~\equiv x,y).
\label{eq:Yukawa-perp}
\end{eqnarray}
Then, if the quadratic term is dominant, 
the negative behavior of the perpendicular-type color-magnetic correlation, 
$C_{\perp}(r) < 0$, could be explained. 
However, the real situation is not so simple.

\begin{figure}[htbp]
    \centering
    \includegraphics[width=7.5cm]{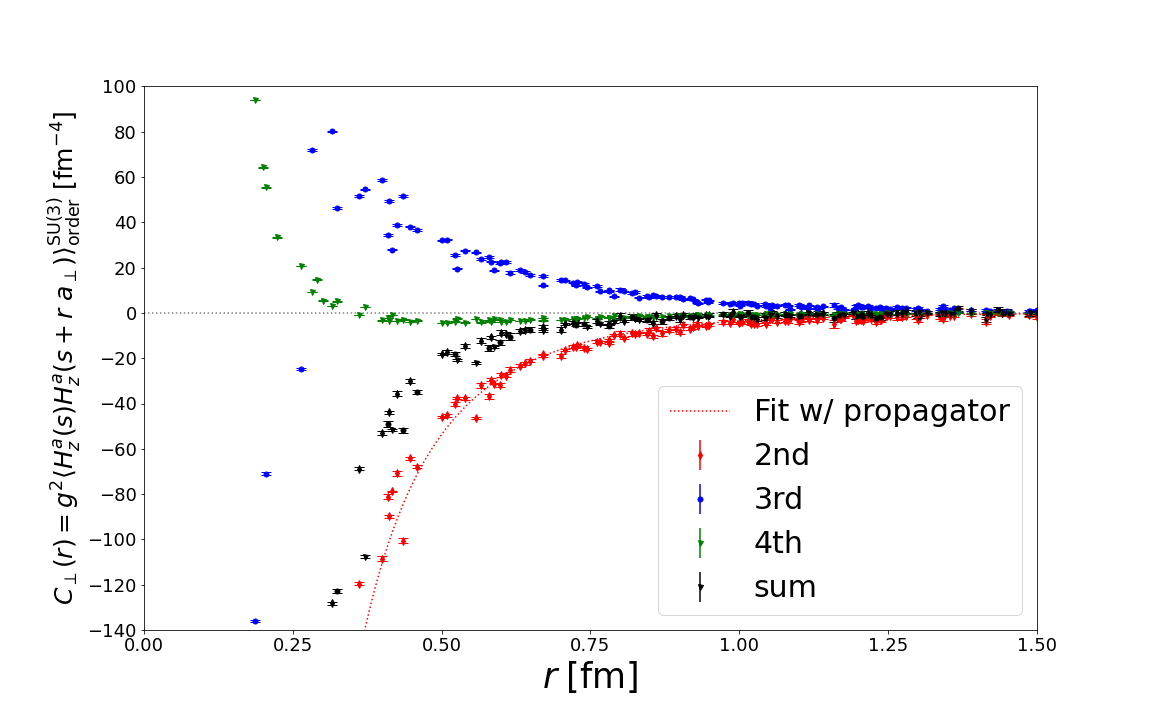}
    \includegraphics[width=7.5cm]{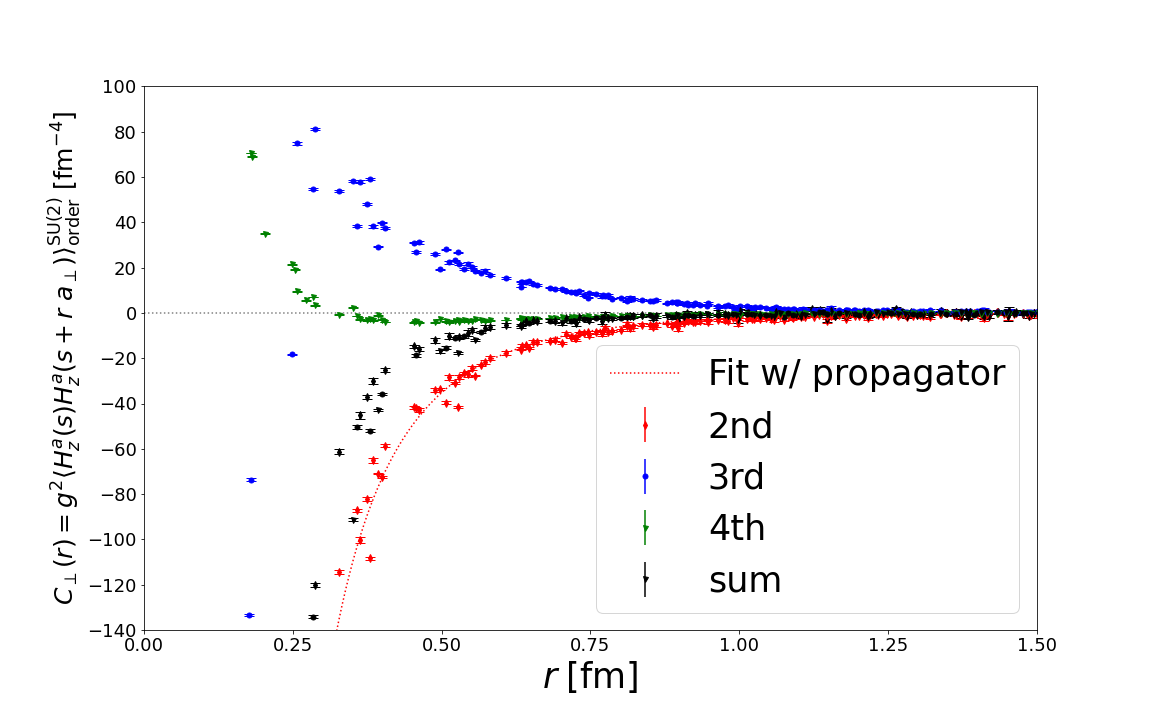}
    \caption{
Each contribution of 
the quadratic (red), cubic (blue) and quartic (green)
terms of the perpendicular-type color-magnetic correlation $C_{\perp}(r)$ (black)
in the Landau gauge in SU(3) (left) and SU(2) (right) lattice QCD.
The red dotted line denotes the curve of Eq.~(\ref{eq:Yukawa-perp}) derived from the Yukawa-type propagator 
$D_{\rm Yukawa}(r)$. 
    }
    \label{fig:decomposition}
\end{figure}

Figure~\ref{fig:decomposition} 
shows individual contributions of 
the quadratic, cubic and quartic 
terms in the perpendicular-type color-magnetic correlation $C_{\perp}(r)$ in lattice QCD.
The quadratic term is always negative,
as was demonstrated with the Yukawa-type gluon propagator in the Landau gauge. 
The cubic term is comparable to the quadratic term, whereas the quartic term gives a relatively small contribution.
In the infrared region, 
the cubic term is positive 
and tends to cancel with 
the quadratic term, 
resulting in a small value of $C_{\perp}(r)$.
%
%
Since the cubic term of the gauge field is unique to non-
abelian gauge theories, its significant contribution in the
QCD vacuum indicates the distinction between QCD and
abelian gauge theories.

\section{Summary and Conclusion}

To examine color-magnetic instabilities in QCD, we have studied field-strength correlations in both SU(2) and SU(3) lattice QCD.
In the Euclidean Landau gauge, 
we have numerically calculated 
the perpendicular-type color-magnetic correlation, 
$C_{\perp}(r) \equiv g^2 \langle H_z^a(s)H_z^a(s + r\hat \perp)) \rangle$ ($\perp \equiv x, y$), 
and the parallel-type one,  
$C_{\parallel}(r) \equiv g^2 \langle H_z^a(s)H_z^a(s + r\hat \parallel) \rangle$ ($\parallel~\equiv z, t$). 

Curiously, we have found that the perpendicular-type color-magnetic correlation 
$C_{\perp}(r)$ is always negative for 
all values of $r$, except for $r=0$. 
In contrast, we have found that 
the parallel-type color-magnetic correlation $C_{\parallel}(r)$
is always positive.
In the infrared region, 
$C_{\perp}(r)$ and $C_{\parallel}(r)$ 
strongly cancel each other, 
which leads to an approximate cancellation 
for the sum of the field-strength correlations as 
$\sum_{\mu, \nu} \langle G^a_{\mu\nu}(s)G^a_{\mu\nu}(s')\rangle 
\propto C_{\perp}(|s-s'|)+ C_{\parallel}(|s-s'|) 
\simeq 0$.

Next, we have decomposed 
the perpendicular-type color-magnetic correlation $C_{\perp}(r)$ into 
the quadratic, cubic and quartic terms of the gluon field $A_\mu$.
The quadratic term is always negative, which can be explained with the Yukawa-type gluon propagator in the Landau gauge. The quartic term gives a relatively small contribution. 
In the infrared region, the cubic term is positive and tends to cancel
with the quadratic term, resulting in  
small $C_{\perp}(r)$.

Finally, we consider the negativity of the perpendicular-type color-magnetic correlation, $C_{\perp}(r)<0$.
If it were an abelian gauge theory, 
this could be be explained by the magnetic-flux conservation, but such an argument cannot be applied to QCD.
%
%
The negative correlation seems to contradict 
the simple constant-magnetic or multi-vortex picture. 
Instead, the negative correlation $C_{\perp}(r)<0$ indicates that color-magnetic fields are 
highly stochastic in the QCD vacuum
\cite{NO79,D87S88,DS88,GP92,EGM97,BBV98}. 

\section*{Acknowledgment}

We thank Prof. 
Jeff Greensite for his useful comments. 
H.S. is supported in part 
by the Grants-in-Aid for
Scientific Research [19K03869] from JSPS.
K.T. is supported by SPRING Program at Kyoto University. 
The lattice QCD calculation has been performed by SQUID at Osaka University.

\end{document}